\documentclass{svproc}
\usepackage[utf8]{inputenc}

\usepackage{url}

\usepackage{natbib}
\bibpunct{(}{)}{;}{a}{}{,}%
\usepackage{graphicx}
\usepackage{amsmath,amssymb}
\usepackage{multirow}
\usepackage{longtable}
\usepackage{changes}
\usepackage{xcolor} 
\usepackage{floatflt}         

\usepackage[b5paper]{geometry}
\def\apj{ApJ}

\def\apjs{ApJ Suppl. Ser.}

\def\mnras{MNRAS}

\def\lum{\rm erg~s$^{-1}$}

\newcommand{\msun}{$M_\odot$}

\newcommand{\kms}{\:\mbox{km~s$^{-1}$}}
\newcommand{\ace}{\mbox {$\alpha_{ce}$}}
\def\lum{\rm erg~s$^{-1}$}
\def\apgt{\ {\raise-.5ex\hbox{$\buildrel>\over\sim$}}\ }
\def\aplt{\ {\raise-.5ex\hbox{$\buildrel<\over\sim$}}\ }

\begin{document}
	\mainmatter              
	\title{X-ray luminosity function of accreting neutron stars and black holes}
	\titlerunning{X-ray luminosity function}  
	%
	\author{Konstantin Postnov\inst{1} 
\and Alexandre Kuranov\inst{1} 
\and Lev Yungelson\inst{2}
\and Marat Gil'fanov\inst{3}}
	\authorrunning{Konstantin Postnov et al.} 
	%
	\tocauthor{Konstantin Postnov, Alexandre Kuranov, Lev Yungelson, Marat Gilfanov}
	\institute{Sternberg Astronomical Institute, 14 Universitetskij pr., Moscow, 
Russia\\
 \email{kpostnov@gmail.com}\\
\and
Institute of Astronomy, RAS,  48 Pyatnitskaya str., Moscow, Russia
\and
Space Research Institute, RAS, 84/32 Profsoyuznaya str., Moscow, Russia}

		\maketitle 
\begin{abstract}
We model X-ray luminosity functions (XLF) of
accreting neutron stars and black holes in $10^{35} \leq L_X \leq 10^{41}$\, \lum\ range in star-forming galaxies and galaxies with the initial star formation burst.  
XLFs are obtained by combining a fast
generation of compact object+normal star population using the binary population synthesis code BSE and  
calculation 
of the subsequent detailed binary evolution  by the MESA code. 
XLF in the galaxies of both types is broadly reproduced using the standard assumptions of the binary star evolution.

	\keywords{neutron stars, black holes, X-ray sources}
\end{abstract}

\section{Introduction}
\label{sec:intro}
Stellar-mass field X-ray sources in galaxies are usually identified with binaries harbouring neutron stars (NS) or black holes (BH) accreting matter from their companions.
Here, we consider the X-ray luminosity 
functions (XLF) of model galaxies with an instantaneous star-formation burst 
as a proxy for elliptical galaxies and a model with constant star-formation rate (SFR) as a
proxy for star-forming galaxies. In \S\ref{sec:method}, we present our assumptions and method of computations, in \S\ref{sec:results} we show the results of computations and discuss them.

\section{Method of Computations}
\label{sec:method}
The matter from the optical star in a binary system with NS or BH can be transferred to the compact object in different ways -- via the Roche lobe overflow (RLOF) of the donor star or by the gravitational capture of the stellar wind from the optical companion.
In the case of RLOF onto magnetized NSs, we take into account that the NS magnetic fields of  
about $10^{12} \aplt B \aplt 10^{14}$\,G modify the standard Shakura-Sunyaev disk accretion pattern. This can result in a 
super-Eddington  X-ray luminosity $L_X$ from the accreting NS even before the  Eddington accretion rate is
attained \citep{2020AstL...46..658K}. This effect allows us to explain the existence of 
ultra-luminous X-ray sources (ULX) with NS without assuming a 
strong beaming of X-ray radiation, which is confirmed by the inconsistency of the strong beaming with the observed high pulsed fraction of emission from pulsating ULXs \citep{2021MNRAS.501.2424M}. In the case of wind accretion, we include in the model the subsonic settling quasi-spherical accretion onto magnetized NS which is efficient if 
$\dot{M} \aplt 4\times 10^{16}$\,g\,s$^{-1}$ \citep{2012MNRAS.420..216S,2015ARep...59..645S}. Otherwise, both BH and NS experience supersonic Bondi-Hoyle-Lyttleton accretion. We also take into account transiency of LMXB with unstable accretion disks  \citep{1999MNRAS.303..139D}. 

To model the population of X-ray binaries, we applied a "hybrid" method:
the evolution of binaries up to the formation of compact object (c.o.) 
is computed by an updated and modified rapid 
binary population synthesis (BPS) code BSE 
\citep{htp02}, while their further detailed evolution with mass transfer onto the c.o. is followed by the interpolation in the  grid
of evolutionary tracks precomputed by a regular stellar evolution code MESA
\citep[][and references therein]{2015ApJS..220...15P}. 

\begin{floatingfigure}[l]{0.46\textwidth}
\hspace{-.28\textwidth}
\vspace{-1.5cm}
		\includegraphics[width=0.9\textwidth]{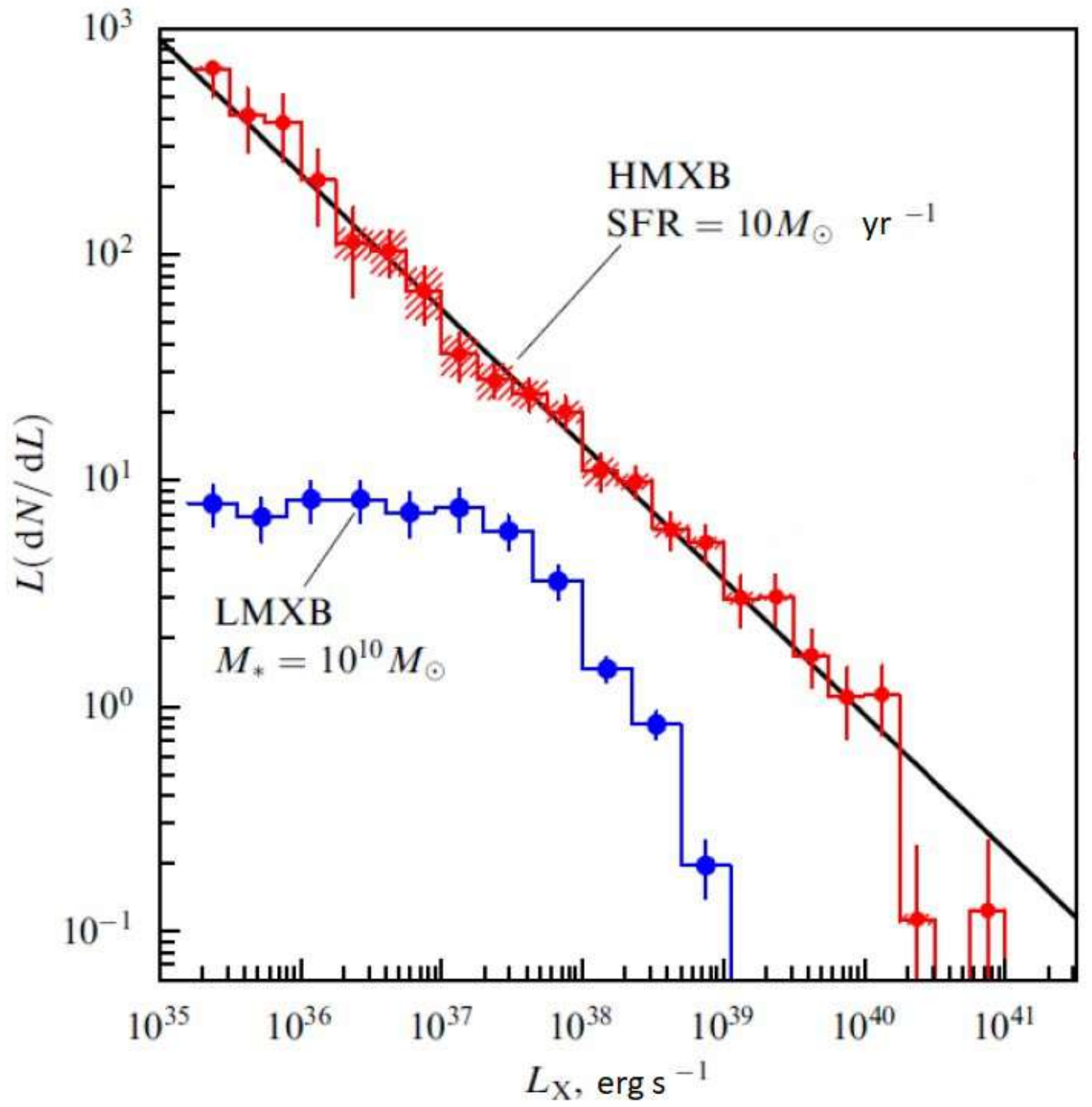}
		\caption{XLF of accreting binaries in star-forming and elliptical galaxies, HMXB and LMXB, respectively \citep{2013PhyU...56..714G}.}
\label{fig:obs}
\end{floatingfigure}

In the BPS computations, 
the common envelope stage is treated using the 
$\alpha\lambda$ formalism 
\citep{web84,dek90} with the ``common envelope efficiency'' \ace=1 and the 
binding energy parameter $\lambda$\  from \citet{2011ApJ...743...49L}.
Among the most pressing issues in stellar evolution is the formation mechanism of NSs and BHs. It is still an unsolved 3D-problem. Ignoring the details of this process, as common in BPS, we adopt one of the  1D-approximations discussed in the literature; namely,
we assume that the mass of c.o. is equal to 90\% of the pre-SN CO core mass. 
A unique mass 1.4\,\msun\ is assigned to all newborn NSs. The kick velocity 
received by nascent NS is assumed to follow a Maxwellian distribution with $\sigma$=265\\\kms\ 
\citep{hobbs2005}. An exception are NSs produced by the electron-capture SN (ECSN)  from  (8 -- 9)\,\msun\ progenitors, 
to which we assigned (quite arbitrarily) Maxwellian kicks with $\sigma$=30\,\kms. Similar kicks reduced by the factor $M_{BH}/M_{OV}$, where $M_{OV}=2.5 M_\odot$ is the assumed  maximum possible NS mass, are 
assigned to nascent BHs. 
A more detailed description of our  assumptions and method can be found in 
\citep{2020MNRAS.496L...6Y,2020AstL...46..658K}.

\section{Results and Discussion}
\label{sec:results}

\begin{figure*}[t!] 
\hspace{-0.53cm}
\begin{minipage}[t]{0.49\textwidth}
\includegraphics[scale=0.28]{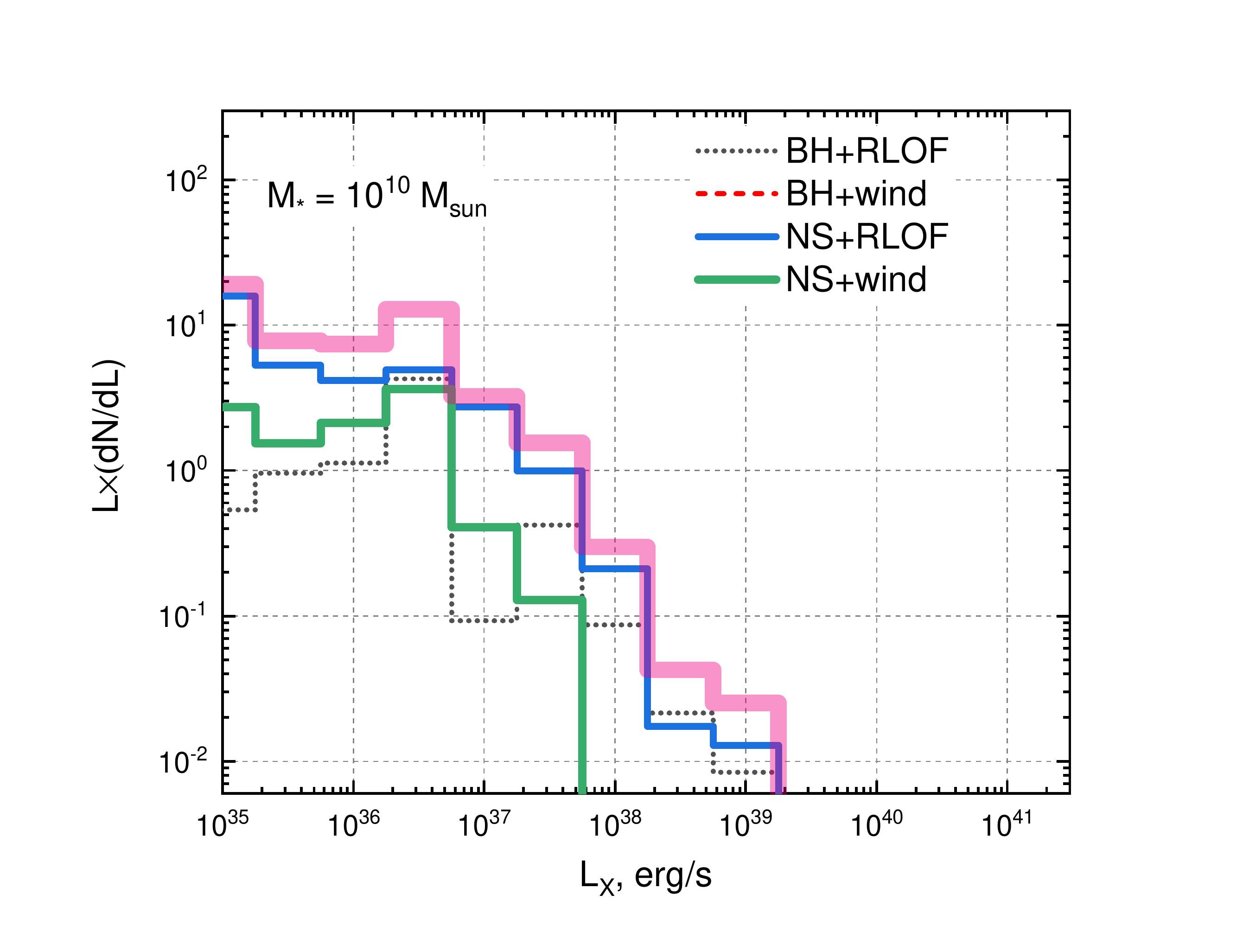}
\hskip 0.02\textwidth
\end{minipage}
\begin{minipage}[t]{0.49\textwidth}
\includegraphics[scale=0.28]{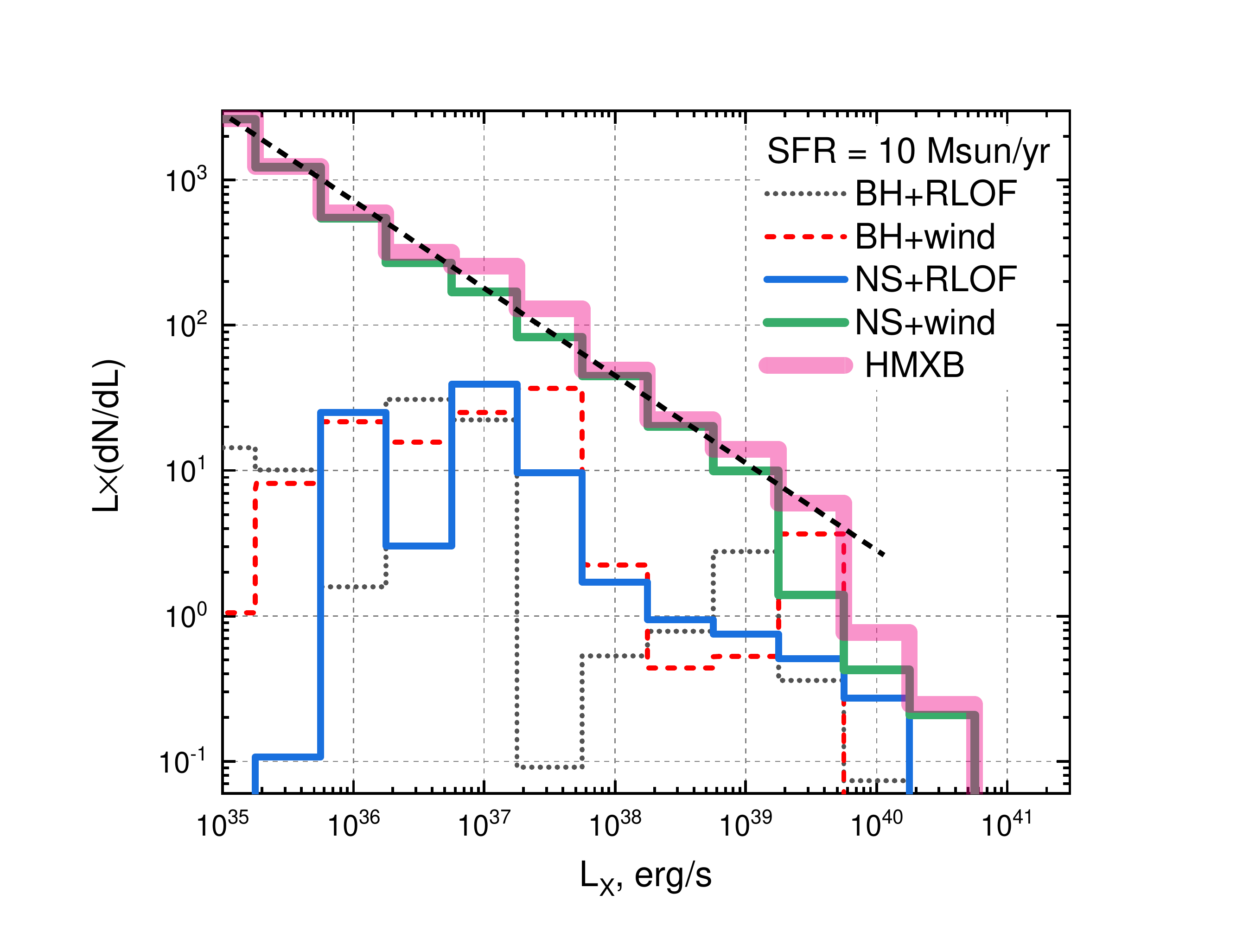}
\end{minipage}
\vspace{-1cm}
\caption{XLF for a $M_\star = 10^{10}$\,\msun\  model galaxy with instantaneous burst of star formation (left panel) and star-forming galaxy of the same mass with constant SFR (right panel) at t=10\,Gyr.  Separately is shown the contribution of sources with different c.o. and accretion patterns. Thick magenta line shows the total XLF. Dashed line indicates the slope (-0.6) of XLF for HMXB. }
\label{fig:mod}
\end{figure*}  

It is known that in the absence of a substantial emission from the central supermassive object, the X-ray luminosity of star-forming galaxies is dominated by high-mass X-ray binaries (HMXB) with initial donor masses  exceeding (8 -- 10)\,\msun, while in elliptical galaxies  the binaries with the initial donor masses $\sim 1$\,\msun\ (LMXB)  dominate. This allowed one to conclude that, owing to the difference in the evolutionary timescales, in 
the star-forming galaxies the number of X-ray sources should be proportional to the star formation rate (SFR), while in the ellipticals --- to the total mass of the galaxy $M_\star$ (see, e.g., \citet{2013PhyU...56..714G} and references therein). XLF (defined as $L_X\times \frac{dN}{dL_X}$) for the samples of galaxies of two types is shown in Fig.~\ref{fig:obs}. The lines in this plot can be compared
to the model XLF for galaxies with similar mass ($10^{10}$\,\msun),  in which either constant continuous SFR or instantaneous burst of star formation were assumed (Fig.~\ref{fig:mod}). Crudely, they may be taken as proxies for star-forming and elliptical galaxies, respectively. 

Remarkably, the simplest assumptions on the c.o. masses 
and moderate kicks for BH and post-ECSN NSs allowed us to reproduce the slope of the observed XLF for HMXB close to -0.6. The broken shape of XLF for ``ellipticals'' reflects the ``death'' of systems with high and 
moderate mass donors soon after the star formation  burst. Only stars with 
$\sim 1$\,\msun\ donors  survived after $\sim 10$\,Gyr\ of evolution. In ``spirals'', massive binaries producing HMXBs 
are born continuously and, simultaneously, binaries with low- and moderate-mass donors born at earlier epochs replenish the population of X-ray sources. 
Note that in ``spirals'', XLF is always dominated by the systems with NS accreting from stellar wind, thanks to the permanent presence of sufficiently massive stars prior to RLOF (while
the latter stage is, predominantly, unstable). In ``ellipticals'', at late stages, giant donor stars with strong winds have already finished their evolution. In this case, low-mass semidetached systems, in which the donors experienced RLOF on the main-sequence or shortly after TAMS and have low mass-loss rates, mainly contribute to XLF. 

X-ray sources with $L_X \apgt 10^{39}$\,\lum\ appropriate to the  Eddington luminosity of 
$\apgt$10\,\msun\ BH  are classified as ULX. 
Figure~\ref{fig:mod} shows
that in  the populations of both types the objects with NS accretors should dominate. Non-detection of them in elliptical galaxies may suggest that the NS magnetic fields decayed or were buried by accretion long ago, preventing their appearance as ULX. 

\vskip 0.3cm 
The authors acknowledge support by RFBR grant 19-12-00790. L.Y. was partially supported by RFBR grant 19-07-01198. The work of K.P. and A.K. is supported by the Interdisciplinary Scientific Educational School of Moscow University 'Fundamental and applied space research'.
\vskip 1cm
\begingroup              
\let\clearpage\relax

\endgroup
\end{document}